\documentstyle[epsf]{l-aa}
\newcommand{\eq}[1]{Eq.\,\ref{#1}}
\newcommand{\tab}[1]{Tab.\,\ref{#1}}
\newcommand{\fig}[1]{Fig.\,\ref{#1}}
\newcommand{\msun}{M_{\sun}} 
\newcommand{\mzams}{M_{\rm ZAMS}}
\newcommand{\dup}{dredge-up}
\newcommand{\cdr}{$^{13}$C}
\newcommand{\hevi}{$^{4}$He}
\newcommand{\czw}{$^{12}$C}
\newcommand{\ose}{$^{16}$O}
\newcommand{\nezw}{$^{22}$Ne}
\newcommand{\mgfu}{$^{25}$Mg}
\begin{document}
\thesaurus{06         
          (08.05.3;   
           08.16.4;   
           08.01.1)   
          }
\title{Stellar evolution of low and intermediate-mass stars}

\subtitle{IV.\, 
Hydrodynamically-based overshoot and nucleosynthesis in AGB stars}

\author{F.\ Herwig\inst{1}, T.\ Bl\"ocker\inst{2}, D. Sch\"onberner\inst{1} 
        and M. El Eid\inst{3}}

\offprints{F.\ Herwig}

\institute{
  Astrophysikalisches Institut Potsdam, 14773 Potsdam,
  Germany (fherwig@aip.de; deschoenberner@aip.de)
\and
  Institut f\"ur Astronomie und Astrophysik, Universit\"at Kiel,
  24098 Kiel, Germany (bloecker@astrophysik.uni-kiel.de)
\and
  American University of Beirut, Department of Physics, Beirut, Lebanon
  (meid@aub.edu.lb)
}

\date{Received date; accepted date}
\maketitle

\begin{abstract}
The focus of this study is on the 
treatment of those stellar regions immediately adjacent
to convective zones. 
The results of hydrodynamical simulations by Freytag et al.\ (1996)
show that 
the motion of convective elements extends well beyond the
boundary of the convectively unstable region.
We have applied their parametrized description of the
corresponding velocities 
to the treatment of overshoot in stellar evolution calculations
up to the AGB (Pop.\,I, $\mzams=3\msun$).

Our calculations show the
3$^{\rm rd}$ \dup\ already at the $7^{\rm th}$ thermal pulse (TP),
and the \dup\ parameter reaches \hbox{$\lambda = 0.6$}
during the next five pulses.
Accordingly, the amount of  dredged up \czw\ 
is up to $10^{-3} \msun$. Our models develope a small so-called \cdr-pocket 
consisting of a few $10^{-7} \msun$.
Finally, this treatment of boundaries of convective regions leads
to intershell abundances of typically  (\hevi/\czw/\ose)=(23/50/25) 
(compared to (70/26/1) in the standard treatment).
\end{abstract}

\section{Introduction}
The question of burning and mixing along the AGB is intimately linked to the
so-called {\it carbon star mystery} (Iben 1981): Observations of the LMC and SMC
prove that most of the carbon (and s-process element) enriched AGB stars are
found at rather low luminosities indicating that they also have rather 
{\it low} masses (1...$3 M_{\odot}$,  
cf.\ Smith et al.\ 1987, Frogel et al.\ 1990). On the other hand, AGB models
predict the  3$^{\rm rd}$ \dup, i.e.\ the mixing of interior carbon to the 
surface during the recurrent He shell flashes on the AGB, mostly for much
higher luminosities, i.e. larger (core) masses. Thus, we meet two problems:
(i) Why do we not observe luminous carbon stars?; and (ii) Why do we not 
find dredge up in low mass AGB models? 

The first one belongs most likely to the 
occurrence of hot bottom burning and mass loss 
(Iben 1975, Bl\"ocker \& Sch\"onberner 1991, Boothroyd et al.\ 1993), see 
also D'Antona \& Mazzitelli (1996) for a recent discussion.

The second problem will be subject of this {\it Letter}. 
Synthetic AGB calculations  utilize the 
stellar parameters known from evolutionary AGB model sequences and take
the minimum core mass for dredge up, $M_{\rm H}^{\rm min}$, and the 
dredge-up parameter $\lambda$ (ratio of dredged-up mass to  growth of
the hydrogen exhausted core per flash cycle)
 as adjustable parameters (among others) 
in order to match the observations, 
i.e. the luminosity function of LMC carbon stars.
For example, van~den Hoek \& Groenewegen (1997) find 
$M_{\rm H}^{\rm min} = 0.58 $M$_{\odot}$ and $\lambda=0.75$ as best
fit in contrast to the results of evolutionary calculations 
($M_{\rm H}^{\rm min} \ga 0.65 $M$_{\odot}$ and  $\lambda \approx 0.25$,
see Wood 1997). The amount of dredge-up found in evolutionary calculations
depends  sensitively on metallicities, core and 
envelope masses of the models (Wood 1981).
Additionally, as pointed out by 
Frost \& Lattanzio (1996), numerical details may play an important role.
Finally, it is in particular the treatment of convection 
which bears  large uncertainties concerning the dregde-up efficiency.
 
Another problem related to this subject concerns the \mbox{s-process}
nucleosynthesis. The required neutrons are probably
not released via the  \nezw($\alpha,n$)\mgfu\  reaction.
Instead, the s-process is most likely driven by the \cdr($\alpha,n$)\ose\
neutron source, raising the question how to mix protons into carbon-rich
layers in order to produce sufficient amounts of \cdr\ . 
Iben \& Renzini (1982) found that the  protons can diffuse
from the bottom of the convective envelope into the intershell zone due to
semiconvection. However, this scenario is restricted to low metallicities 
(Iben 1983). Often, ``standard calculations'' 
artificially  ingest a given amount of \cdr\ at the onset of the pulse
into the convectively unstable He shell. However, Straniero et al. (1995) have 
shown that \cdr\ formed during the dredge-up phase is burnt under radiative
conditions during the interpulse phase (for a recent review see Lattanzio 1995).

These difficulties of standard stellar evolution calculations
to confront the observations 
often led to the conclusion that
mixing  outside the formally convective boundaries may take place 
(e.g. Hollowell \& Iben 1988, D'Antona \& Mazzitelli 1996, 
Wood 1997),
or even that only a hydrodynamic approach of modelling the H/He interface 
will overcome the drawbacks of the local treatment of convection 
(Arlandini et al.\ 1995).

In brief the situation can be summarized as follows:
The mixing length theory (MLT) (B\"ohm-Vitense, 1958) is usually applied 
to stellar evolution calculations.
But it does not describe additional (possibly partial)
mixing beyond the Schwarzschild boundary.
To meet the observations for main sequence stars the 
{\em instantaneous}
mixing of the convective core is extended  
beyond the classical Schwarzschild boundary by
 some fraction of the pressure scale height.
This treatment is commonly referred to as overshoot. It
has been discussed for example by 
Alongi et al.\ (1993, and ref. therein).
The method we present in the following 
has in some aspects similar consequences as
former overshoot treatments
but it is based on hydrodynamic calculations.

Freytag et al.\
(1996) have carried out  two-dimensional  numerical radiation 
hydrodynamics simulations to study the structure and 
dynamics of a variety of shallow surface convection 
zones. Their work 
reveals an independent theoretical evidence
for the actual existence of extra mixing beyond the 
boundary of convectively unstable regions.
The parametrisation of the exponentially declining 
velocities of convective elements beyond the classical convective 
border can be  applied to one-dimensional stellar evolution 
calculations. This treatment leads then to some extra partial 
mixing (Bl\"ocker et al.\ 1997).
In the following chapters we confine ourselves to
the description of the method and the most important consequences.
Further implications and details will be given elsewhere.

\section{The stellar evolution code and the method of mixing}
The computations for this study are based on the code
described by Bl\"ocker (1993, 1995) with major modifications.
Nuclear burning is accounted for 
via a nucleosynthesis network including 31 isotopes and 74 reactions
up to carbon burning. We use the most recent opacities (Iglesias
et al. 1992, Alexander \& Ferguson 1994). The initial composition is
$(Y,Z)=(0.28,0.02)$,  the mixing length parameter
of the MLT is $\alpha = 1.7$. 
For the $3 M_{\odot}$ model sequence presented in this paper 
a Reimers-type mass loss with $\eta=1$ has been applied.

\subsection{Equation for abundance changes}
The main modification relevant for this study is the introduction
of a time-dependent model for mixing in the formulation adopted
by Langer et al.\ (1985) for the study of semiconvection in 
massive stars:
\begin{equation}
        \frac{{\rm d}X_i}{{\rm d}t}=\left( \frac{\partial{X_i}}{\partial t} 
        \right)
        _{\rm nuc} + \frac{\partial}{\partial M_{\rm r}}
        \left[ \left(4\pi r^2\rho\right)^2 D 
        \frac{\partial X_i}{\partial M_{\rm r}} \right] \label{diff-gl}
\end{equation}
with $X_i$ being the mass fraction of the respective isotope. 
The first term 
on the right-hand side considers the abundance changes due to nuclear 
burning and the corresponding rates are determined by the 
nucleosynthesis network. 
The second part describes the 
mixing of the elements by means of a diffusion 
term given here with respect to the mass coordinate 
$M_{\rm r}$. 
The diffusion coefficient $D$ depends on the assumed mixing model.
For instance, in convective unstable regions
within the Schwarzschild boundaries $D$ is 
derived from the MLT (see below).

In our present calculations we
make the assumption that the burning rates, i.e.\ 
$(\partial{X_i}/\partial t)_{\rm nuc}$, do
not change during one time step. 
The nuclear network is invoked just once for each time
step and the resulting burning rates enter \eq{diff-gl}.

\subsection{The choice of D}
\eq{diff-gl} is integrated  over the whole star 
for each isotope. 
For the radiative zones $D$ is obviously  $D_r=0$.
Within  the  convectively unstable regions
we follow the prescription of 
Langer et al.\ (1985): $D_{\rm c}=1/3v_{\rm c}l$ with 
$l$ being the mixing
length and $v_{\rm c}$ the average velocity of the 
convective elements according to the MLT
(B\"ohm-Vitense, 1958). Then the diffusion
coefficient can be written as
\begin{equation}
     D_{\rm c}= \frac{1}{3} \alpha ^ {2/3} H_{\rm p}
     \left[ \frac{c}{\kappa\rho} g \beta (1- \beta)
     \nabla_{\rm ad}(\nabla_{\rm rad} - \nabla_{\rm ad}) \right] ^{1/3},
\end{equation}
where $H_{\rm p}$ is the pressure scale height, $\kappa$ the opacity,
$\alpha$ the mixing length parameter, c the velocity of light, 
$\rho$ the density, g the gravitational acceleration, 
$\nabla_{\rm rad}$ and $\nabla_{\rm ad}$ the radiative and adiabatic
temperature gradient respectively, and $\beta$ the gas pressure fraction. 

Finally, for the regions 
which are immediately
adjacent to convectively unstable zones we adopt 
the depth dependent diffusion coefficient derived 
by Freytag et al.\ (1996: Eq.\,9) from their numerical simulations of
two-dimensional radiation hydrodynamics of time-dependent 
compressible convection :
\begin{equation} \label{dhyd}
           D_{\rm os} = D_0 \exp{\frac{-2 z}{H_{\rm v}}}, \,\,\,\,\,\,\,
           D_0 = v_{0} \cdot H_{\rm p}, \,\,\,\,\,\,\,\,
           H_{\rm v} = f \cdot H_{\rm p},  
\end{equation}
where $z$ denotes the distance from the edge of the convective zone 
($z=|r_{\rm edge}-r|$ with $r$: radius), and 
$H_{\rm v}$ is the velocity scale height 
of the overshooting convective elements at $r_{\rm edge}$ 
being  proportional to the pressure scale height $H_{\rm p}$. 
According to Freytag et al.\ (1996), $D_{0}$ has to be taken near the 
edge of the convective zone. Here the velocity field varies only slighty.
We take the corresponding typical velocity of the convective elements, 
$v_{0}$, obtained from the MLT in the unstable layers immediately before the 
Schwarzschild border. 

\begin{figure}
\centering
\epsfxsize=0.48\textwidth
\mbox{\epsffile{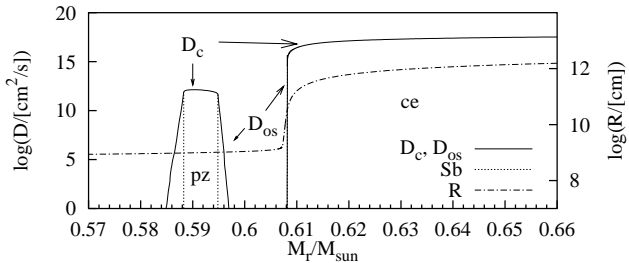}}    
\caption[]{Diffusion coefficient ($D_{\rm c}$, $D_{\rm os}$) versus mass
coordinate for an TP-AGB model with $f=0.02$, $49 {\rm yr}$ before the
largest \hevi-luminosity (beginning of $12^{\rm th}$ pulse cycle). The
pulse-driven convective zone (pz) occupies the lower half of  
the intershell zone. The latter extends from $0.587\msun$ (location
of the \hevi-shell) to
$0.607\msun$ (where H-burning takes place).
The vertical dotted lines indicate the Schwarzschild boundaries (Sb). For the
base ($M_{\rm r}=0.608\msun$) of the convective envelope (ce) 
it can not be  distinguished from $D_{\rm os}$ since the
H-shell is very extended. This can be seen from the dashed-dotted line 
which shows the radius (r). However, this overshoot region comprises
$5.8*10^{-5}\msun$ accounted for in 16 mass shells.} \label{diff-ms}
\end{figure}
We summarize the choice of $D$ (\fig{diff-ms}):
\begin{displaymath}
D=
\left\{
       \begin{array}{lp{6.5cm}}
      D_{\rm c}  & {\rm instantaneous mixing, ordinary convection}\\
      D_{\rm os} & {\rm mixing efficiency declining, overshoot region}\\
      D_{\rm r}  & {\rm no mixing, radiative layers}
       \end{array}
       \right.
\end{displaymath}

\subsection{The determination of $f$}
The parameter $f$ is a measure of the efficiency of the extra partial mixing. 
It defines the velocity scale height of
the convective elements beyond the boundary of the convectively unstable
zone. As can be seen from \eq{dhyd} the decline of $D_{\rm os}$
is steeper for smaller values of $f$, 
thus the larger $f$ the
further extends the extra partial mixing beyond the convective edge.

Freytag et al.\ (1996) find
from their simulations of overadiabatic convective envelopes of A-stars and
DA white dwarfs $f=0.25 \pm 0.05$ and $1.0 \pm 0.1$ respectively. 
These results already reveal that $f$ 
depends on the stellar parameters.
For deep envelope (or
core) convection $f$ can be expected to be considerably smaller since the
ratio of the Brunt-V\"ais\"al\"a timescales of the 
stable to the instable
layers decrease with increasing depth, i.e. adiabacity. 
However, hydrodynamical
simulations of such deep convection zones are not available yet
and we have to rely on indirect methods. 
For example, $f \approx 0.02$ results in the same main-sequence width
as calculations by Schaller et~al.\ (1992) who applied instantaneous
overshoot with a parameter of $d_{\rm over}/H_{\rm p}=0.2$.
We took this value of $f$ for our calculations assuming that it holds also
for the deep convection zones of  AGB stars.
Although
the width of the calculated main sequence depends very sensitively on $f$
($\Delta f = 0.005$ corresponds to a displacement of the terminating age
main sequence of $\Delta \log{g} \approx 0.08$) our experiments with
AGB evolution calculations have shown that the character of our
findings remains qualitatively valid even if $f$ is changed by a factor
of two. 

\section{First results from the calculations}
Applying this treatment of convection to AGB models we found 
the $3^{\rm rd}$ dredge up for a $3 M_{\odot}$ model. The onset was already 
at the $7^{\rm th}$ thermal pulse at $M_{\rm H} = 0.58 M_{\odot}$
reaching a dredge-up parameter of $\lambda = 0.6$ within the next five pulses
(see Tab.~1). Note, that corresponding calculations without extra mixing 
(standard calculations) did not show any dredge up.

Due to the application of depth-dependent overshoot to all convective zones the
abundances in the intershell zone change in  comparison to
standard calculations. The pulse-driven convective zone extends deeper
into the underlying C/O-core already during the first TP and
considerable amounts of \ose\ and \czw\ are mixed up.
Compared to the values given in \tab{DUP-TAB}
the abundances of the intershell zone in standard calculations is 
typically (\hevi/\czw/\ose)=(70/26/1). 
The comparison of these intershell abundances with observed surface abundances
of Wolf-Rayet (WR) central stars 
and PG\,1159 stars can be used in the future as an additional
constraint for $f$.
                       
\begin{table}
\caption[iso]{Dredge-up data for selected thermal pulses of a $M=3 \msun$
sequence: mass of the hydrogen exhausted core ($M_{\rm H}$), 
\dup\ parameter
($\lambda = \Delta M_{\rm Dup} / \Delta M_{\rm H}$, 
$\Delta M_{\rm H}$: increase  of the core mass by nuclear burning
during the last pulse cycle),
total amount of dredged-up material ($M_{\rm DUP}$),
amount of dredged up  \czw\ 
($M_{{\rm DUP}(^{12}{\rm C})}$),
composition of the intershell zone, 
 and content of the \cdr-pocket 
($M_{(^{13}{\rm C})}$).
} \label{DUP-TAB} 
\begin{flushleft}
\begin{tabular}{lllll} 
\hline \noalign{\smallskip}
TP Nr.                                  & & $7$           & $8$           & $12$  \\ 
\noalign{\smallskip}
\hline      
$\frac{M_{\rm H}}{\msun}$                &  & $0.5779$      & $0.5837$      & $0.6073$\\
$\lambda$                              &  & $0.11$        & $0.12$        & $0.60$  \\
$\frac{M_{\rm DUP}}{\msun} $            &     & $6.3*10^{-4}$ & $7.0*10^{-4}$ & $3.7*10^{-3}$ \\
$\frac{M_{{\rm DUP}(^{12}{\rm C})}}{\msun} $  & & $1.4*10^{-4}$ & $1.8*10^{-4}$ & $1.5*10^{-3}$ \\
(\hevi/\czw/\ose)                      &  & $(23/50/25)$  & $(23/49/26)$  & $(26/45/27)$ \\
$\frac{M_{(^{13}{\rm C})}}{\msun} $    &  & $3.9*10^{-7}$ & $3.7*10^{-7}$ & $2.0*10^{-7}$ \\ \hline
\end{tabular}
\end{flushleft}
\end{table}
\begin{figure}
\centering
\epsfxsize=0.48\textwidth
\mbox{\epsffile{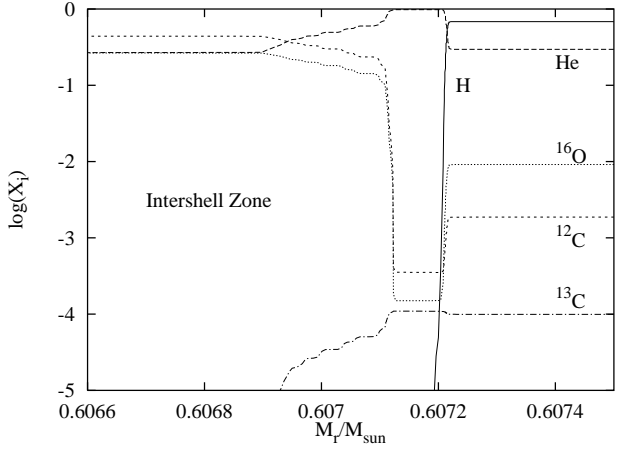}}    
\caption[]{Onset of the 3rd \dup\ during the 12$^{\rm th}$ pulse cycle. 
 At $0.6072 \msun$ is the bottom of the convective envelope which has already
 moved into the \hevi-dominant layer  at $M_{r}/M_{\odot}\approx
 0.6071-0.6072$. The latter  is the product of the H-shell burning of the
 previous pulse. This situation is located $273 {\rm yr}$ after
 the beginning of $12^{\rm th}$ pulse cycle, i.e. $322 {\rm yr}$ later than
 shown in \fig{diff-ms}.}
 \label{DUP-start}
\end{figure}
\begin{figure}
\centering
\epsfxsize=0.48\textwidth
\mbox{\epsffile{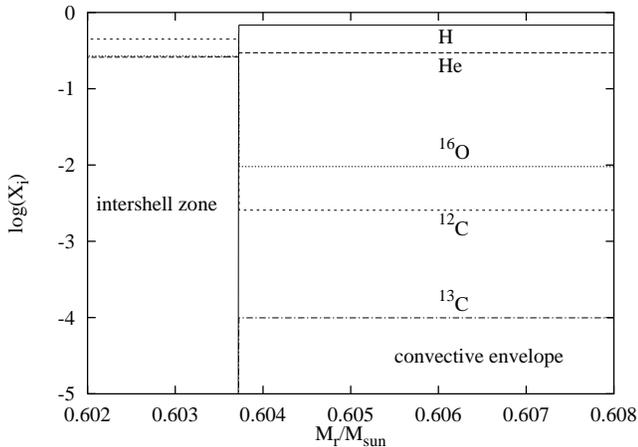}}    
\caption[]{The convective envelope continues to proceed
downwards ($210 {\rm yr}$ past \fig{DUP-start}) and will shortly hereafter
stop.  
Note the
different mass range compared to the previous figure.} \label{DUP-continue}
\end{figure}
\begin{figure}
\centering
\epsfxsize=0.48\textwidth
\mbox{\epsffile{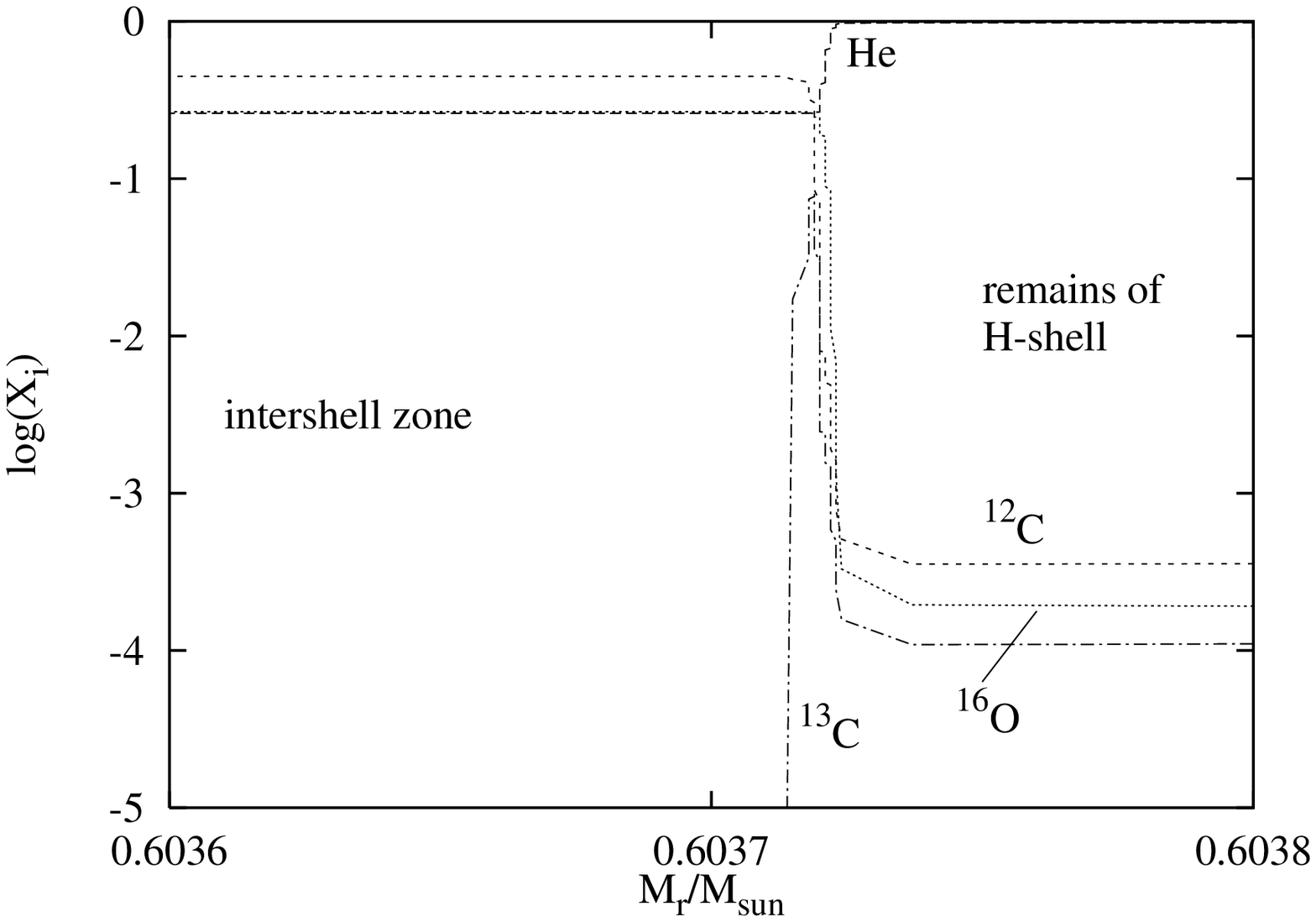}}
\caption[]{The downward movement of the 
convective envelope has stopped, the \dup\ is over. 
 A small \cdr-pocket has formed in the region of
previously overlaping diffusion tails of th H- and \czw-profile.
The snapshot is $17660 {\rm yr}$ later than the \fig{DUP-continue}.}
\label{DUP-final}
\end{figure}

This strong increase of \czw\ in particular causes now a corresponding
increase of opacity and thereby of the radiative gradient 
$\nabla_{\rm rad}$  and leads to semiconvection (Iben \& Renzini 1982)
immediately below the \hevi-rich layer. 
In our calculations this shows
up as thin convective shells which may be connected by
their overlaping diffusive tails. The effect is that the sharp border
between the intershell zone and the \hevi-rich layer is wiped out
(\fig{DUP-start}, $M \approx 0.607\msun$). 
In the following, the envelope convection continuously progresses 
downwards, and finally its base proceeds
through the previously semiconvective zone (\fig{DUP-continue}). 
Typically, 50\% of the dredged-up material consist of $^{12}$C, the rest is 
given by $^{4}$He and $^{16}$O, resp., with 25\% each (\tab{DUP-TAB}).

Finally the convective envelope base remains 
at the position of deepest
penetration for about fifty years.
This is the phase where the adjacent H and
\czw\ profiles can built up a very thin overlap zone due to the diffusive
tail of the envelope convection ($D_{\rm os}$, \eq{dhyd}).
At this location in the star the \cdr-pocket can form (\fig{DUP-final}).
We confirm the findings of Straniero et al. (1995) that the built up \cdr- 
pocket is burnt under  radiative conditions well before the next thermal pulse.

\section{Concluding remarks}
We considered overshoot in evolutionary calculations based on  results 
of 2-dimensional hydrodynamical simulations of convection. On the AGB,
this leads to 
the formation of a \cdr-pocket in the intershell region 
in contrast to calculations based on standard overshoot considerations 
(i.e. on  formal shifts of the Schwarz\-schild border). 
Additionally, the intershell abundances are considerably changed due to 
extra mixing.
As opposed to calculations
without any overshoot,
we find now  dregde-up  even for models with masses
as low as $3 $M$_{\odot}$.
%
\begin{acknowledgements}
We are grateful to F.\,Rogers and  D.\,Alexander 
for providing us with their opacity tables.
We thank 
M.\,Steffen and B.\ Freytag for valuable discussions about 
hydrodynamic simulations of  convection.
F.H.\,and T.B.\,acknowledge funding by the Deutsche Forschungsgemeinschaft
(grants Scho\,394/13 and We\,1312/10-1).
\end{acknowledgements}


\begin{thebibliography}{}

\bibitem{} Alexander D.R., Ferguson J.W., 1994, ApJ 437, 879. 

\bibitem{} Alongi M., Bertelli G., Bressan A., Chiosi C., Fagotto F.,
Greggio L., Nasi E., 1993, A\&A, 97 851.

\bibitem{} Arlandini C., Gallino R., Busso M., Straniero O., 1995, in 
 {\em 32${\rm nd}$ Li\`ege Int. Astrophys. Coll.}, 
 eds. A. Noels et al., p.\,447. 

\bibitem{} Bl\"ocker T., 1993, Ph.D. thesis, University of Kiel

\bibitem{} Bl\"ocker T., 1995, A\&A 297, 727.

\bibitem{} Bl\"ocker T., Herwig F., Sch\"onberner D., El Eid M., 1997,
in {\em The Carbon Star Phenomenon}, IAU Symp.\ 177, in press.

\bibitem{} Bl\"ocker T., Sch\"onberner, D. 1991, A\&A 244, L43.

\bibitem{} B\"ohm-Vitense E., 1958, Z. Astrophys. 46, 108.

\bibitem{} Boothroyd A.D., Sackmann I-J., Ahern S.C., 1993, ApJ 416, 762

\bibitem{} D'Antona F., Mazzitelli I., 1996, ApJ 470, 1093.

\bibitem{} Freytag B., Ludwig H.-G., Steffen M., 1996, A\&A 313, 497.

\bibitem{} Frost C.A., Lattanzio J.C., 1996, ApJ 473, 383.

\bibitem{} Frogel J.A., Mould J.R., Blanco V.M, 1990, ApJ 352, 96.

\bibitem{} van~den Hoek L.B., Groenewegen M.A.T., 1997, A\&A, in press.

\bibitem{} Hollowell D., Iben I. Jr., 1988, ApJ 333, L25.

\bibitem{} Iben I. Jr., 1981, ApJ 246, 278.

\bibitem{} Iben I. Jr., 1983, ApJ 275, 65.

\bibitem{} Iben I. Jr., 1975, ApJ 196, 525.

\bibitem{} Iben I. Jr., Renzini A., 1982, ApJ 263, L23.

\bibitem{} Iglesias C.A., Rogers F.J., Wilson B., 1992, ApJ 397, 717.

\bibitem{} Langer N., El Eid M., Fricke K.J., 1985, A\&A 145, 179.

\bibitem{} Lattanzio J.C., 1995, in {\em Nuclei in the Cosmos},
  eds. M. Busso, R. Gallino, C.M. Raitieri, AIP Conf.\ Ser.\, 327, p.\ 353.

\bibitem{} Schaller G., Schaerer D., Meynet G., Maeder A., 1992, A\&A 96,
269.

\bibitem{} Smith V.V., Lambert D.L., McWilliam A., 1987, ApJ 320, 862.

\bibitem{} Straniero O., Gallino R., Busso M., Chieffi A., Raitieri C.M., 
Salaris M., Limongi M., 1995, ApJ 440, L85.

\bibitem{} Wood P.R., 1981, in  
  {\em Physical Processes in Red Giants}, eds. I. Iben, A. Renzini, Reidel,
  Dordrecht, p. 135

\bibitem{} Wood P.R., 1997, in  
  {\it Planetary Nebulae}, IAU Symp.\, 180, Kluwer, Dordrecht, in press

\end{thebibliography}
\end{document}